\documentclass[%
 reprint,
 superscriptaddress,
 preprintnumbers,
 bibnotes,
 amsmath,amssymb,
 aps,
]{revtex4-1}

\usepackage[utf8]{inputenc}
\usepackage[dvips]{graphicx}
\usepackage{color}
\usepackage{array}
\usepackage{slashed}
\usepackage{overpic}
\usepackage{tikz}
\usepackage{hyperref}
\usepackage[caption=false]{subfig}
\usepackage{arydshln}
\usepackage{comment}
\usepackage{yhmath}
\usetikzlibrary{shapes.geometric,arrows,decorations.pathmorphing,decorations.markings,patterns}

\usepackage{color}

\newif\ifdraft
\drafttrue

\def\spa#1.#2{\left\langle#1\,#2\right\rangle}
\def\spb#1.#2{\left[#1\,#2\right]}

\newcommand{\eq}{\begin{equation}}
\newcommand{\eqe}{\end{equation}}
\newcommand{\eqa}{\begin{eqnarray}}
\newcommand{\eqae}{\end{eqnarray}}

\newcommand{\bea}{\begin{eqnarray}}
\newcommand{\eea}{\end{eqnarray}}

\newbox\charbox
\newbox\slabox
\def\s#1{{      
        \setbox\charbox=\hbox{$#1$}
        \setbox\slabox=\hbox{$/$}
        \dimen\charbox=\ht\slabox
        \advance\dimen\charbox by -\dp\slabox
        \advance\dimen\charbox by -\ht\charbox
        \advance\dimen\charbox by \dp\charbox
        \divide\dimen\charbox by 2
        \raise-\dimen\charbox\hbox to \wd\charbox{\hss/\hss}
        \llap{$#1$}
}}

\def\be{\begin{equation}}
\def\ee{\end{equation}}
\def\ba{\begin{eqnarray}}
\def\ea{\end{eqnarray}}

\def\CP1{\mathbb{CP}^1}
\def\SL2C{\mathrm{SL}(2,\mathbb{C})}

\def\Z2{\mathbb{Z}_2}

\def\su2{{SU(2)}}

\def\[{\left[}
\def\]{\right]}

\def\s{\sigma}

\def\({\left(}
\def\){\right)}
\def\[{\left[}
\def\]{\right]}

\def\<{\langle}
\def\>{\rangle}

\def\i2{\frac{i}{2}}

\def\tu{{\tilde u}}

\def\2F1{\,_2{\rm F}_1}

\definecolor{mygreen}{rgb}{0,0.4,0}

\begin{document}

\title{
Binary Geometries, Generalized Particles and Strings, and Cluster Algebras
} 
\author{Nima Arkani-Hamed}%
 \affiliation{School of Natural Sciences, Institute for Advanced Studies, Princeton, NJ, 08540, USA \\
Center of Mathematical Sciences and Applications, Harvard University, Cambridge, MA 02138, USA}
\author{Song He}
\affiliation{%
CAS Key Laboratory of Theoretical Physics, Institute of Theoretical Physics, Chinese Academy of Sciences, Beijing 100190, China\\
School of Physical Sciences, University of Chinese Academy of Sciences, No.19A Yuquan Road, Beijing 100049, China}
\author{Thomas Lam}
\affiliation{Department of Mathematics, University of Michigan, 530 Church St, Ann Arbor, MI 48109, USA \\
Department of Mathematics, Massachusetts Institute of Technology, 77 Massachusetts Ave., Cambridge, MA 02139, USA}
\author{Hugh Thomas}
\affiliation{LaCIM, D\'epartement de Math\'ematiques, Universit\'e du Qu\'ebec \`a Montr\'eal, Montr\'eal, QC, Canada}

\date{\today}

\begin{abstract}
We introduce the notion of ``binary" positive and complex geometries, giving a completely rigid geometric realization of the combinatorics of generalized associahedra attached to any Dynkin diagram. We also define open and closed ``cluster string integrals" associated with these ``cluster configuration spaces". The binary geometry of type ${\cal A}$ gives a gauge-invariant description of the usual open and closed string moduli spaces for tree scattering, making no explicit reference to a worldsheet. The binary geometries and cluster string integrals for other Dynkin types provide a generalization of particle and string scattering amplitudes. Both the binary geometries and cluster string integrals enjoy remarkable factorization properties at finite $\alpha'$, obtained simply by removing nodes of the Dynkin diagram. As $\alpha'\to 0$ these cluster string integrals reduce to the canonical forms of the ABHY generalized associahedron polytopes.  For classical Dynkin types these are associated with $n$-particle scattering in the bi-adjoint  $\phi^3$ theory through one-loop order.
\end{abstract}

\maketitle


\date{\today}

\section{Binary Geometries}
A remarkable fact about particle scattering in arbitrary spacetime dimension is that the poset of planar singularities form combinatorial polytopes. The associahedron polytope~\cite{Stasheff_1, Stasheff_2} encodes combinatorially all planar cubic tree graphs, which has been realized directly in the kinematic space and known as the ABHY {\it kinematic associahedron}~\cite{Arkani-Hamed:2017mur}. Its canonical form~\cite{Arkani-Hamed:2017tmz} computes the tree-level S-matrix of bi-adjoint $\phi^3$-theory, which makes hidden symmetries of the amplitudes manifest.  In~\cite{bazier2018abhy} the ABHY realization was extended to {\it generalized associahedra of any finite type cluster algebra}~\cite{fomin2003systems, chapoton2002polytopal}, whose canonical forms compute bi-adjoint $\phi^3$-amplitudes through one loop~\cite{Arkani-Hamed:2019vag}; the canonical form of the type ${\cal B}$ or ${\cal C}$ polytope (known as the cyclohedron) contains one-loop tadpole diagrams, while for the type ${\cal D}$ polytope it gives the integrand for one-loop bi-adjoint $\phi^3$-amplitudes. 

For a finite type, rank-$n$ cluster algebra, these are $n$-dimensional polytopes whose facets (resp., vertices) are in one-to-one correspondence with cluster variables (resp., cluster seeds). The boundary structure of such a polytope has interesting factorizations encoded in the corresponding Dynkin diagram, where each facet corresponds to removing a node of the Dynkin diagram. For the associahedron of type ${\cal A}_n$, the facets factorize as ${\cal A}_m \times {\cal A}_{n{-}1{-}m}$; the type-${\cal B}_n/{\cal C}_n$ associahedron has facets of the shape ${\cal B}_m \times {\cal A}_{n{-}1{-}m}$ or ${\cal C}_m \times {\cal A}_{n{-}1{-}m}$; the type-${\cal D}_n$ associahedron has facets of the shape ${\cal D}_m \times {\cal A}_{n{-}1{-}m}$, ${\cal A}_1 \times {\cal A}_1 \times {\cal A}_{n{-}3}$ or simply ${\cal A}_{n{-}1}$. Examples of these factorizations are shown below.
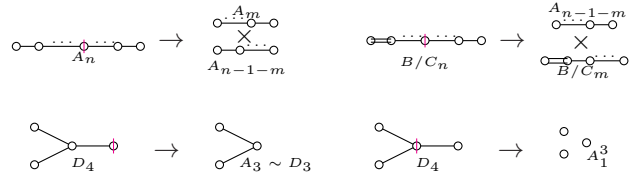
\vspace{-1em}
\begin{center}
\begin{tikzpicture}[baseline={([yshift=0.5ex]current bounding box.center)},scale=0.3]
\draw (2.5,0) -- (-3,0);
\node[fill=white,circle,draw=black, inner sep=0pt,minimum size=3pt] at (2.5,0) {};
\node[fill=white,circle,draw=black, inner sep=0pt,minimum size=3pt] at (1.5,0) {};
\node[fill=white,circle,draw=black, inner sep=0pt,minimum size=3pt] at (0,0) {};
\node[fill=white,circle,draw=black, inner sep=0pt,minimum size=3pt] at (-3,0) {};
\node[fill=white,circle,draw=black, inner sep=0pt,minimum size=3pt] at (-2,0) {};
\draw[magenta] (0,-0.3) -- (0,0.3);
\node at (1,0.2) {\tiny \dots};
\node at (-0.9,0.2) {\tiny \dots};
\node at (0,-0.5) {{\tiny $A_n$}};
\end{tikzpicture}
$\, \to \,$
\begin{tikzpicture}[baseline={([yshift=-.5ex]current bounding box.center)},scale=0.3]
\draw (2.5,0) -- (0,0);
\node[fill=white,circle,draw=black, inner sep=0pt,minimum size=3pt] at (2.5,0) {};
\node[fill=white,circle,draw=black, inner sep=0pt,minimum size=3pt] at (1.5,0) {};
\node[fill=white,circle,draw=black, inner sep=0pt,minimum size=3pt] at (0,0) {};
\node at (0.8,0.2) {\tiny \dots};
\draw (2.5,-1.2) -- (0,-1.2);
\node[fill=white,circle,draw=black, inner sep=0pt,minimum size=3pt] at (2.5,-1.2) {};
\node[fill=white,circle,draw=black, inner sep=0pt,minimum size=3pt] at (1,-1.2) {};
\node[fill=white,circle,draw=black, inner sep=0pt,minimum size=3pt] at (0,-1.2) {};
\node at (1.8,-1) {\tiny \dots};
\node at (1.2,-0.6) {$\times$};
\node at (1.3,0.5) {\tiny $A_m$};
\node at (1.3,-2) {\tiny $A_{n{-}1{-}m}$}; 
\end{tikzpicture}
$\qquad$
    \begin{tikzpicture}[baseline={([yshift=0.5ex]current bounding box.center)},scale=0.3]
            \draw (2.5,0) -- (-1.5,0);
            \node[fill=white,circle,draw=black, inner sep=0pt,minimum size=3pt] at (2.5,0) {};
            \node[fill=white,circle,draw=black, inner sep=0pt,minimum size=3pt] at (1.5,0) {};
            \node[fill=white,circle,draw=black, inner sep=0pt,minimum size=3pt] at (0,0) {};
            \node[fill=white,circle,draw=black, inner sep=0pt,minimum size=3pt] at (-2.4,0) {};
            \node[fill=white,circle,draw=black, inner sep=0pt,minimum size=3pt] at (-1.5,0) {};
            \draw[magenta] (0,-0.3) -- (0,0.3);
            \node at (1,0.2) {\tiny \dots};
            \node at (-0.5,0.2) {\tiny \dots};
            \draw (-1.5,-0.07) -- (-2.4,-0.07);
            \draw (-1.5,0.07) -- (-2.4,0.07);
            \node at (0,-1) {\tiny $B/C_n$};
    \end{tikzpicture}
    $\, \to \,$ 
    \begin{tikzpicture}[baseline={([yshift=-0.5ex]current bounding box.center)},scale=0.3]
            \draw (2.5,0) -- (0,0);
            \node[fill=white,circle,draw=black, inner sep=0pt,minimum size=3pt] at (2.5,0) {};
            \node[fill=white,circle,draw=black, inner sep=0pt,minimum size=3pt] at (1.5,0) {};
            \node[fill=white,circle,draw=black, inner sep=0pt,minimum size=3pt] at (0,0) {};
            \node at (0.8,0.2) {\tiny \dots};
            \draw (3,-1.6) -- (0.5,-1.6);
            \node[fill=white,circle,draw=black, inner sep=0pt,minimum size=3pt] at (3,-1.6) {};
            \node[fill=white,circle,draw=black, inner sep=0pt,minimum size=3pt] at (1.5,-1.6) {};
            \node[fill=white,circle,draw=black, inner sep=0pt,minimum size=3pt] at (0.5,-1.6) {};
            \draw (0.5,-1.7) -- (-0.5,-1.7);
            \draw (0.5,-1.5) -- (-0.5,-1.5);
            \node at (2.2,-1.4) {\tiny \dots};
            \node[fill=white,circle,draw=black, inner sep=0pt,minimum size=3pt] at (-0.5,-1.6) {};
            \node at (1.1,-0.8) {$\times$};
            \node at (1.2,-2.1) {\tiny $B/C_m$};
            \node at (1.4,0.6) {\tiny $A_{n{-}1{-}m}$};
    \end{tikzpicture}  
\end{center}
\begin{center}
    \begin{tikzpicture}[baseline={([yshift=-.5ex]current bounding box.center)},scale=0.5]
            \draw (-0.9,0) -- (-2,0);
            \draw (-2,0) -- (-3,0.5);
            \draw (-2,0) -- (-3,-0.5);
            \node[fill=white,circle,draw=black, inner sep=0pt,minimum size=3pt] at (-0.9,0) {};
            \node[fill=white,circle,draw=black, inner sep=0pt,minimum size=3pt] at (-3,-0.5) {};
            \node[fill=white,circle,draw=black, inner sep=0pt,minimum size=3pt] at (-2,0) {};
            \node[fill=white,circle,draw=black, inner sep=0pt,minimum size=3pt] at (-3,0.5) {};
            \draw[magenta] (-0.9,-0.2) -- (-0.9,0.2);
            \node at (-1.7,-0.5) {\tiny $D_4$};
    \end{tikzpicture}
    $\quad \to \quad$ 
    \begin{tikzpicture}[baseline={([yshift=-.5ex]current bounding box.center)},scale=0.5]
            \draw (-2,0) -- (-3,0.5);
            \draw (-2,0) -- (-3,-0.5);
            \node[fill=white,circle,draw=black, inner sep=0pt,minimum size=3pt] at (-3,-0.5) {};
            \node[fill=white,circle,draw=black, inner sep=0pt,minimum size=3pt] at (-2,0) {};
            \node[fill=white,circle,draw=black, inner sep=0pt,minimum size=3pt] at (-3,0.5) {};
            \node at (-1.5,-0.5) {\tiny $A_3\sim D_3$};
    \end{tikzpicture}\qquad
    \begin{tikzpicture}[baseline={([yshift=-.5ex]current bounding box.center)},scale=0.5]
            \draw (-0.9,0) -- (-2,0);
            \draw (-2,0) -- (-3,0.5);
            \draw (-2,0) -- (-3,-0.5);
            \node[fill=white,circle,draw=black, inner sep=0pt,minimum size=3pt] at (-0.9,0) {};
            \node[fill=white,circle,draw=black, inner sep=0pt,minimum size=3pt] at (-3,-0.5) {};
            \node[fill=white,circle,draw=black, inner sep=0pt,minimum size=3pt] at (-2,0) {};
            \node[fill=white,circle,draw=black, inner sep=0pt,minimum size=3pt] at (-3,0.5) {};
            \draw[magenta] (-2,0.2) -- (-2,-0.2);
            \node at (-1.7,-0.5) {\tiny $D_4$};
    \end{tikzpicture}
    $\quad \to \quad$
    \begin{tikzpicture}[baseline={([yshift=-.5ex]current bounding box.center)},scale=0.3]
            \node[fill=white,circle,draw=black, inner sep=0pt,minimum size=3pt] at (-2,0) {};
            \node[fill=white,circle,draw=black, inner sep=0pt,minimum size=3pt] at (-3,-0.5) {};
            \node[fill=white,circle,draw=black, inner sep=0pt,minimum size=3pt] at (-3,0.5) {};
            \node at (-1.5,-0.6) {\tiny $A_1^3$};
    \end{tikzpicture}
\end{center} 

It is fascinating that cluster polytopes provide a geometric realization of the combinatorics of ``compatibility" for cluster variables. Each cluster variable is associated with a facet of the polytope; compatible variables correspond to facets in the polytope that meet, while incompatible variables correspond to facets that do not touch at all. But there is still freedom in the particular realization of the polytope, and any realization can be continuously deformed in various ways without changing the relevant combinatorics. It is therefore natural to wonder whether any geometric realization of the combinatorics exists that is more rigid and canonical. 

This is what we seek to do in this letter. We introduce the idea of {\bf binary geometries} as a canonical and rigid way of realizing the combinatorics of generalized associahedra for any finite-type cluster algebra. For each facet or cluster variable denoted as $a$, we assign a $u_a$-variable, and impose the same number of constraints of the form
\be\label{ueqs}
u_a+ \prod_{{\rm all}~b} u_b^{b ||  a}=1\,, \quad \forall a
\ee
where the non-negative integer $b ||  a$ is called the {\it compatibility degree} from $b$ to $a$ (originally defined in~\cite{fomin2003systems}). It is zero if and only if $a$ and $b$ are compatible, or equivalently if the two facets meet. It is highly non-trivial that for $n$-dimensional generalized associahedra, these {\it $u$-equations} are consistent and the solution space turns out to be $n$-dimensional. Demanding $u_a \geq 0$, these equations also force all the $u_a$ to take values in $[0,1]$. Thus when a facet $a$ is reached with $u_a \to 0$, the $u_b$ variables for all {\it incompatible} facets ($b || a>0$) are forced to go to $1$. This is why we call this a ``binary" realization of  generalized associahedra. Restricting the $u$-variables to be positive, the solution space gives a ``curvy" realization of the polytope. It is striking that the ``binary" property is not restricted to real and positive $u_a$, but is a feature of the complex space of solutions of equation \eqref{ueqs}. When $u_a \to 0$, the second term in the equation $u_b+ \prod_{{\rm all}~c} u_c^{c ||  b}=1$ becomes 0 for all incompatible $u_b$'s, and so those $u_b$ all equal to $1$.  This is a novel feature of the binary geometry: almost all the connections between geometries with factorizing boundary structures and physics, have involved reality and positivity in a crucial way, whereas here the factorization holds with complex variables.

For type ${\cal A}$, the binary realization (with $b||a=0,1$ only) turns out to give a gauge-invariant description of the moduli space of open- and closed-string worldsheet at genus zero~\cite{DM} (with $u$-variables being cross-ratios~\cite{Brown:2009qja}). The binary positive and complex associahedron geometry directly underpin open- and closed-string amplitudes; we will see that both the Parke-Taylor form and Koba-Nielsen factor~\cite{Koba:1969kh} in string integrals are naturally written in terms of the $u$-variables. String amplitudes are the {\bf stringy canonical form}, or $\alpha'$-extension of the canonical form, for ABHY associahedra~\cite{Arkani-Hamed:2019mrd}:
not only does the $\alpha'\to 0$ limit reproduce the $\phi^3$-tree amplitude, but it factorizes into products of lower amplitudes at finite $\alpha'$.

We will give a concise definition of the compatibility degrees from the ABHY realization~\cite{Arkani-Hamed:2019vag}, and it is remarkable that $u$-equations encode the boundary structures of generalized associahedra from \eqref{ueqs}. Naturally associated with them are the generalized open- and closed-string amplitudes that we call  {\bf cluster string integrals}~\cite{Arkani-Hamed:2019mrd}; the open cluster string integral is the integral over the binary geometry of its canonical form, and is beautifully regulated by including all the factors of the form $u^{\alpha' X}$
\be
{\cal I}(\{X\})=\int_{U^+} \Omega^{(n)}(U^+) \prod_a u_a^{\alpha' X_a}\,.
\ee
These are the most perfect examples of stringy canonical forms with leading order given by ABHY polytope~\cite{Arkani-Hamed:2019mrd}, and their factorizations at finite $\alpha'$ again correspond to removing a node of the Dynkin diagram~\footnote{As  will be discussed in~\cite{20201}, for non-simply laced case we actually obtain the generalized associahedron of the dual Dynkin diagram.}! For example, the $\alpha'\to 0$ limit of type ${\cal D}$ integral gives one-loop planar $\phi^3$-amplitude, and it factorizes as product of lower-point integrals of type ${\cal D}$ and ${\cal A}$. 

Our aim in this letter is to to give a self-contained description of the $u$-equations, and summarize and highlight only a few of the important features of the binary geometry and associated stringy canonical forms.  We will present a much more detailed discussion of these in a longer companion article \cite{20201}, where we will give a cluster algebraic explanation of $u$-variables and $u$-equations and construct these {\bf cluster configuration spaces} as certain quotients of cluster varieties and show that the space is smooth with normal-crossing boundary divisors.  


\section{Binary associahedra and string amplitudes} 


\paragraph*{The $U$ space and the moduli space.} We begin by discussing the $u$-variables in the most familiar setting of the usual associahedron. We introduce $N:=\frac{n(n{-}3)}2$ $u_{i,j}$-variables, and consider $N$ $u$-equations~\cite{Arkani-Hamed:2017mur}:
\be\label{global}
1-u_{i, j}=\prod_{(k, l)~{\rm cross}~(i, j)} u_{k, l}\,,\quad
\ee
where the product is over all chords $(k, l)$ that cross $(i, j)$, {\it i.e.} all $u$'s that are {\it incompatible} with $u_{i,j}$. 
Eq. \eqref{global} is the type-${\cal A}$ version of \eqref{ueqs} where the only non-zero degree is $1$.  For example for $n=4$ we have $u_{1,3}+u_{2,4}=1$ and for $n=5$, we have $u_{1,3}+u_{2,4} u_{2,5}=1$ and its cyclic images. 

Let's denote the solution space of \eqref{global} as the {\bf $U$ space}, and a non-trivial observation is that $n{-}3$ of the equations are redundant, and thus $U_n$ has dimension $n{-}3$. The $U$ space has the boundary structure of an associahedron purely algebraically: at a boundary $u_{i, j}\to 0$, we see that for any incompatible $(k,l)$, the RHS of $1-u_{k,l}$ vanishes, thus $u_{k, l} \to 1$ and all incompatible $u$'s decouple; we are left with $U$ spaces for two polygons divided by $(i, j)$, thus the boundary is given by their product:
\be\label{Ufactorization}
\partial_{u_{i, j}\to 0} U_n=U(i,i{+}1,\ldots, j) \times U(j, j{+}1, \ldots, i)\,.
\ee
We see that $u$-equations encode the boundary structure of ${\cal A}_{n{-}3}$ even for $u\in \mathbb{C}$. For example, as $u_{1,3}\to 0$, $u_{2,4}, u_{2,5}\to 1$, which decouple from $u$ equations, and we are left with  $u_{1,4}+u_{3,5}=1$ which defines the $n=4$ space $U(3,4,5,1)$ (times the trivial $U(1,2,3)$ space).

We can include positivity: by requiring all $u_{i,j}  \geq 0$, \eqref{global} implies all of them satisfy $0\leq u_{i,j} \leq1$, which cut out a ``curvy" associahedron with $N$ facets; we will refer to it as the {\bf positive part} of $U$ space, $U_n^+$, which has the same shape as ${\cal A}_{n{-}3}$. For $n=5$, the positive part $U_5^+\sim {\cal A}_2$ gives a ``curvy" pentagon~\cite{Arkani-Hamed:2017mur}.

As mentioned above, the $U$ space turns out to provide a SL$(2)$-invariant description of the moduli space (or configuration space), ${\cal M}_{0,n}$, of $n$ points on $\mathbb{P}^1$, where the $u$'s can be interpreted as {\it cross-ratios} of such $n$ points.  Let's see how the moduli space naturally emerges from \eqref{global}. Consider a pair of disjoint sets of points, $A, B$ and we can write $A=\{a{+}1, a{+}2, \cdots, b\}$ and $B=\{c{+}1, c{+}2, \cdots, d\}$ where $a,b,c,d$ are cyclically ordered. We define $U_{A,B}=\prod_{i\in A, j\in B} u_{i,j}$ and note the trivial identity $U_{A, B_1} U_{A, B_2}=U_{A,B}$ for two disjoint sets $B_1 \cup B_2=B$. Furthermore, we define complementary sets $\bar{A}, \bar{B}$: $\bar{A}=\{b{+}1, \cdots, c\}$ and $\bar{B}=\{d{+}1, \cdots, a\}$. 
From \eqref{global} we can deduce that
\be\label{extendglobal}
U_{A,B}+U_{\bar{A}, \bar{B}}=1\,
\ee
Indeed, \eqref{global} is a special case of \eqref{extendglobal}:  choosing $A=\{i\}, B=\{j\}$, the ranges for $k$ and $l$ in \eqref{global} are $\bar{\{i\}}$, $\bar{\{j\}}$, and \eqref{global} becomes $U_{\{i\}, \{j\}}+U_{\bar{\{i\}}, \bar{\{j\}}}=1$.
The {\it extended} $u$-equations \eqref{extendglobal} nicely lead to the identification of $U$'s as cross-ratios. Denoting $U_{A,B}$ as $[a, b | c, d]$, we have $U_{\bar{A}, \bar{B}}=[b, c | d, a]$, and thus we see the appearance of four points and the identities $[a, b| c, d]$ must satisfy. Trivially, we have $[a, b | c, e] [a, b | e, d]=[a, b | c, d]$; more interestingly by \eqref{extendglobal} we have $[a, b | c, d]+[b, c | d, a]=1$. They are precisely identities that invariantly characterize cross-ratios for $n$ points on $\mathbb{P}^1$! The solutions are
 \be\label{crossratio} [a, b | c, d]=\frac{(a d) (b c)}{(a c)(b d)}\,,\ee 
where $(a b)$ denotes a minor of $G (2,n)$ that represents $n$ points on $\mathbb{P}^1$. If we further require the $n$ points to be ordered on $\mathbb{RP}^1$ (equivalently they can be represented by a point in $G_+(2,n)$), then all the cross-ratios are between $0$ and $1$: $0<[a, b | c, d]<1$. This is how the open-string moduli space, ${\cal M}_{0,n}^+$, emerges from $u$-equations! 

We have just seen that the $u_{i,j}$ allow us to describe the open-string moduli space without thinking about $n$ ordered points on the boundary of a disk. But this standard picture allows us to see something else as well: the presence of {\it different orderings} of the $n$ points, which relate to different color orderings for scattering amplitudes. By contrast, the $u_{i,j}$ variables appear inexorably linked to a single ordering. So how can we see the other orderings from this point of view? This question has a natural answer. 
As we have discussed, if we restrict all the $u_{i, j} \geq0$, that the $u$-equations  force them to lie in the unit interval: $0 \leq u_{i, j} \leq 1$. But it is natural to ask whether some of the $u_{i, j}$'s might be negative. Without any detailed study of the solutions of the extended $u$ equations, we can just ask which sign patterns for the $u$'s are allowed compatible with $U_{A,B} + U_{\bar{A},\bar{B}} = 1$, {\it i.e.} excluding only those sign patterns for which both terms are negative. Quite beautifully, we find that precisely $(n{-}1)!/2$ such sign patterns are allowed.  For any consistent sign pattern, it is natural define new positive variables $\hat{u}$ in the obvious way by writing $\hat{u}_{i,j} = (-1)^{{\rm sgn}(u_{i,j})} u_{i,j}$. We can then re-arrange the extended $u$ equations to be in the form of setting the sum of two monomials in the $\hat{u}$ variables to unity. Remarkably, the set of all the $\hat{U}$ equations we get in this way is just a a re-ordering of the extended $u$ equations we started with!  This exposes a hidden $S_n$-symmetry of the $U_n$ space which we can recognize as permutations of the $n$ points once we identify the space with ${\cal M}_{0,n}$.  Taking the $u_{i,j}$ to be real, we see that $U_n (\mathbb{R})\sim {\cal M}_{0,n}(\mathbb{R})$ is tiled by $(n{-}1)!/2$ {\it connected components}, and each of them is an associahedron $U_n^+ \sim {\cal A}_{n{-}3}$ for a cyclic ordering.  For example, it is easy to check that there are $12$ such consistent sign patterns for $n=5$ (for $12$ pentagons of ${\cal M}_{0,5}(\mathbb{R})$); similarly there are $60$ sign patterns for $n=6$, which gives $60$ ${\cal A}_3$ associahedra that tile ${\cal M}_{0,6}(\mathbb{R})$.

\paragraph*{Open string integrals on the $U$ space.} It is very natural to define integrals on the $U$ space, which turn out to be usual  string integrals with Koba-Nielsen factor. For the open-string case, we are interested in integrating the {\it canonical form} for the positive part, which can be obtained by a (trivial) pushforward~\cite{Arkani-Hamed:2017tmz}. Given an acyclic quiver of ${\cal A}_{n{-}3}$, or equivalently a triangulation of an $n$-gon without any internal triangle, it turns out by \eqref{global} that one can solve for all the $N$ $u$-variables {\it rationally} in terms of the $n{-}3$ $u_\alpha$'s in that cluster seed. 
This provides a one-to-one map from the space $\{0<u_\alpha<1\}$ to $U_n^+$, and thus the canonical form can be obtained by pushforward~\cite{Arkani-Hamed:2017tmz}
\be\label{canonicalform}
\Omega(U_n^+)=\prod_\alpha^{n{-}3} d\log \frac{u_\alpha}{1-u_\alpha}
\,.
\ee
and any acyclic quiver gives the same result. Consider integrating it over $U_n^+$, which of course diverges at boundaries; a natural way to regularize the integral is by putting a factor $u^{\alpha' X}$ with $X>0$ for each boundary: 
\be\label{stringintegral}
{\cal I}^{U^+}_n (\{X\}):=(\alpha')^{n{-}3} \int_{U_n^+}~\Omega(U_n^+)~\prod_{i,j} u_{i,j}^{\alpha' X_{i,j}}\,.
\ee
The form $\Omega(U^+_n)$ is the famous Parke-Taylor form for ${\cal M}_{0,n}^+$ (or $G_+(2,n)$ mod torus action): $\Omega(U^+_n)=\frac{d^n z/{\rm SL}(2)}{(12)\cdots (n1)}$, where $z$'s denote the $n$ punctures of ${\cal M}_{0,n}^+$ (or inhomogeneous coordinates of $G_+(2,n)/T$).  As shown in~\cite{Arkani-Hamed:2019mrd}, ${\cal I}^{U^+}_n$ is nothing but the {\bf open-string integral} where the regulator is exactly the Koba-Nielsen factor: $\prod_{i,j} u_{i,j}^{\alpha' X_{i,j}}=\prod_{a,b} (a b)^{\alpha' s_{a,b}}$ with $n \choose 2$ Mandelstam variables given by $s_{a,b}=X_{a,b}+X_{a{+}1, b{+}1}-X_{a, b{+}1}-X_{a{+}1, b}$; since we mod out the torus action, there are $n$ constraints (momentum conservation) $\sum_{b\neq a} s_{a,b}=0$, and thus only $n(n{-}3)/2$ of the $s_{a, b}$ are linearly independent. 

The leading order of ${\cal I}_n^{U^+}$ is given by the bi-adjoint $\phi^3$-amplitude $m_n(\{X\})$, and the same result can be obtained by summing over saddle points~\cite{Arkani-Hamed:2017mur}, known as Cachazo-He-Yuan (CHY) formulas~\cite{Cachazo:2013hca, Cachazo:2013iea}. This connection between $\alpha'\to 0$ limit and ``scattering equations" in the $\alpha'\to \infty$ limit has been understood as a general phenomenon for any stringy canonical form~\cite{Arkani-Hamed:2019mrd} (see also \cite{Mizera:2017rqa}).

What's special about ${\cal I}^{U^+}_n$ is that it factorizes perfectly even at {\it finite} $\alpha'$, which becomes manifest in the form of \eqref{stringintegral}. It is easy to see that ${\cal I}_n^{U^+}$ has a simple pole at each $X_{i,j}=0$, and the residue is given by the integral at the boundary at $u_{i,j}\to 0$; at this boundary we have $\partial_{u_{i,j}\to 0} \Omega(U^+)=\Omega(\partial_{u_{i,j}\to 0} U^+)$ by definition, which factorizes into two lower forms, since the boundary factorizes as \eqref{Ufactorization}. Remarkably, precisely due to \eqref{global}, the Koba-Nielsen factor factorizes accordingly, and we have
\be
{\rm Res}_{X_{i,j}=0}\,{\cal I}_n^{U^+}=\int_{U^+_L \times U^+_R} (\Omega_L \times \Omega_R) \prod_L (u^{\alpha' X}) \times \prod_R (u^{\alpha' X})\,,\nonumber
\ee
which is nothing but ${\cal I}_L \times {\cal I}_R$, where $L$ and $R$ denote the two polygons divided by the chord $(i j)$ in the $n$-gon. 
Let's give an example for $n=5$; ${\cal I}_5^{U^+}$ reads
\be
{\tiny \int_0^1 d\log \frac{u_{1,3}}{1-u_{1,3}}d\log \frac{u_{1,4}}{1-u_{1,4}}u_{1,3}^{X_{1,3}} u_{1,4}^{X_{1,4}} u_{2,4}^{X_{2,4}} u_{2,5}^{X_{2,5}} u_{3,5}^{X_{3,5}}}\,,
\nonumber
\ee 
whose leading order is the canonical function of ${\cal A}_2$ cut out by $5$ facets, $X_{i,j}>0$, with ABHY conditions $c_{1,3}=X_{1,3}+X_{2,4}-X_{1,4}$, $c_{1,4}=X_{1,4}+X_{2,5}-X_{2,4}$, $c_{2,4}=X_{2,4}+X_{3,5}-X_{2,5}$. At finite $\alpha'$, {\it e.g.} as $X_{13}\to 0$, the residue is given by the Veneziano amplitude (times ${\cal I}_3=1$), ${\cal I}_4^{U^+}=\int_0^1 d \log \frac{u}{1-u} u^s (1-u)^t$ with $s=X_{14}$, $t=X_{35}$. 

It is also natural to consider {\bf closed-string integrals} for a pair of orderings, $\alpha$ and $\beta$, in complex $U$ space: 
\be
{\cal I}^{\rm c}_n (\alpha|\beta):=\int_{U(\mathbb{C})} \Omega(U^+_{\alpha}) \prod u^{\alpha' X} \left(\Omega(U^+_{\beta}) \prod u^{\alpha' X} \right)^*\,,\nonumber
\ee
where from the monomial transformation of $u$'s, we see that $\Omega(U^+_\alpha)$ by \eqref{canonicalform} is the Parke-Taylor form of ordering $\alpha$, and the Koba-Nielsen factor is permutation invariant, thus the integral can be written as $\int_{U(\mathbb{C})} \Omega(\alpha) \Omega^*(\beta) |\prod_{a,b} (a b)^{\alpha' s_{a,b}}|^2$. 
Similarly, we can integrate $\Omega(U^+_\alpha)$ in a different region $U^+(\beta)$. The leading order of both integrals are general bi-adjoint $\phi^3$-amplitudes, $m_n(\alpha|\beta)$, given by (the canonical function of) ABHY associahedra with certain facets sent to infinity~\cite{Arkani-Hamed:2019mrd}.

We end with some counting regarding ${\cal M}_{0,n}$. It is well-known that out of all $(n{-}1)!/2$ ordering $\Omega(\alpha)$'s, only $(n{-}2)!$ are linearly independent, which is the number of independent top-dimensional $d\log$ forms, or Parke-Taylor forms~\cite{Kleiss:1988ne}. Moreover, in the presence of the Koba-Nielsen factor, the dimension of the $(n{-}3)$-th (half the dimension of ${\cal M}_{0,n}(\mathbb{C})$) twisted (co-)homology group is $(n{-}3)!$, which is the number of saddle points~\cite{Arkani-Hamed:2019mrd} (see also~\cite{Cachazo:2013gna,Mizera:2017rqa}), thus the matrix ${\cal I}^{\rm c}_n (\alpha|\beta)$ and $m_n(\alpha|\beta)$ has rank $(n{-}3)!$~\cite{Bern:2008qj, BjerrumBohr:2009rd, Stieberger:2009hq, Cachazo:2014xea}. It is known that the Euler characteristic of ${\cal M}_{0,n} (\mathbb{C})$ is given by $(-1)^{n{-}3} (n{-}3)!$~\cite{Arnol'd1969}~\cite{Mizera:2019gea}. 

\section{Cluster configuration spaces}
Now we explain how a binary geometry, called the cluster configuration space, 
can be constructed for generalized associahedra of finite-type cluster algebras. 
The only new ingredient in the $u$-equations, \eqref{ueqs}, is that the compatibility degree $b || a$ can take values other than $0,1$. They form a $N\times N$ matrix for a finite-type cluster algebra with $N$ cluster variables. Let's give a concise definition of the compatibility degree without referring to most of the machinery of cluster algebra, based on a picture of ``walk through all cluster variables" which underpins the ABHY realizations of generalized associahedra~\cite{bazier2018abhy, Arkani-Hamed:2019vag}.
\paragraph*{General $u$ equations from the walk.} 
To define a walk, we start with any {\it acyclic quiver} with $n$ nodes, and at each step pick one of the {\it sources}, {\it i.e.} a node with only outgoing arrows, and reverse all these arrows so they become incoming~\footnote{Each step of the walk is a special kind of {\it mutation}, and it is always possible since the quiver stays {\it acyclic} and has at least one source at each step.}. We assign each node a ``linear" variable $X$ (which corresponds to a facet of the generalized associahedron), and at each step we have exactly one new variable: when mutating (a source) $v$ to $v'$, we replace $X_v$ with $X_{v'}$ and impose a linear relation~\cite{bazier2018abhy, Arkani-Hamed:2019vag}
\be\label{ABHY}
X_v+X_{v'}-\sum_{w \leftarrow v} X_w=c_v\,,
\ee
where $c_v$ is a positive constant labelled by $v$ as well. Note that for quivers that are not simply-laced, there can be integer ``weight" so in general we have $-\sum_w n_w X_w$ instead (with $n_w$ being the weight for $w$ of the arrow $w \leftarrow v$).  We claim that if and only if the quiver is a Dynkin diagram, we can consistently stop the walk and end up with a polytope which is independent of the constants~\cite{Arkani-Hamed:2019vag}. For such a case, we walk $N{-}n$ steps and require all the $N$ variables to be positive, $X>0$, which gives a $n$-dim generalized associahedron with $N$ facets; \eqref{ABHY} allows us to solve all $N$ variables in terms of the original $n$ $X$'s and $N{-}n$ positive $c$'s. It is highly non-trivial that given a Dynkin diagram, starting from any acyclic quiver and walking in any order always give the generalized associahedron (with different ABHY realizations).

For a given variable $b$, we define $b || a$ for all $a$ from a walk with $N{-}n$ steps as follows. Choose any initial quiver where $b$ is a source and for any variable $a$, the degree is defined as the solution to \eqref{ABHY}, $b || a:=X_a$ with the following conditions on initial $X$'s and the $c$'s.  Let's set all the $n$ initial variables $X=0$ (including $X_b=0)$, and we set all $c=0$ except for $c_b=1$ (correspond to the step immediately after $X_b$). By \eqref{ABHY} we see $b||a=X_a$ is an integer ``Green's function" from $b$ to $a$ with the only non-vanishing source from $c_b=1$. 
To obtain the $N\times N$ matrix we need to do such walks starting with all $N$ of the $a$ variables. One can check that in the simply-laced case we always have $a|| b=b|| a$; 
 for type ${\cal A}$ we have only $a||b=0,1$, and for other types we can have $b|| a>1$.
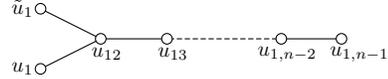
\begin{center}
  \begin{tikzpicture}[baseline={([yshift=-.5ex]current bounding box.center)},scale=0.8,every node/.style={scale=0.8}]    \draw (-1,0) -- (-2,0);    \draw (-2,0) -- (-3,0.5);    \draw (-2,0) -- (-3,-0.5);    \draw [dash pattern={on 2pt off 1pt } ](-1,0) -- (1,0);    \draw (1,0) -- (2,0);    \node[fill=white,circle,draw=black, inner sep=0pt,minimum size=5pt] at (-0.9,0) {};    \node[fill=white,circle,draw=black, inner sep=0pt,minimum size=5pt] at (-3,-0.5) {};    \node[fill=white,circle,draw=black, inner sep=0pt,minimum size=5pt] at (-2,0) {};    \node[fill=white,circle,draw=black, inner sep=0pt,minimum size=5pt] at (-3,0.5) {};    \node[fill=white,circle,draw=black, inner sep=0pt,minimum size=5pt] at (1,0) {};    \node[fill=white,circle,draw=black, inner sep=0pt,minimum size=5pt] at (2,0) {};  \node at (-3.3,-0.5) {$u_1$};    \node at (-3.3,0.5) {$\tilde{u}_1$};    \node at (-1.9,-0.25) {$u_{12}$};    \node at (-0.8,-0.25) {$u_{13}$};    \node at (1.1,-0.25) {$u_{1,n{-}2}$};    \node at (2.3,-0.25) {$u_{1,n{-}1}$};
  \end{tikzpicture}
\end{center}

Let's give some examples of $u$ equations from our definition of compatibility degrees. We start with ${\cal D}_n$ with $N=n^2$ variables, and a natural way to label these variables is to first assign for the Dynkin diagram $n$ initial variables $u_1, \tilde{u}_1, u_{1,2}, u_{1,3}, \cdots, u_{1, n{-}1}$, and the rest can be obtained by cyclic rotations. 
There are three types of $u$ equations, those for $u_i$ or $\tilde{u}_i$, those for $u_{i, i{+}1}$ and those for $u_{i,j}$ for non-adjacent $i,j$. It suffices to write them explicitly for the first non-trivial example, ${\cal D}_4$: 
\ba
\begin{split}\label{D4}
\textrm{4 eqs}: \quad &&1-u_{1,2}&=u_3 \tu_3 u_4 \tu_4 u_{3,4}^2 u_{2,3} u_{2,4} u_{4,1} u_{3,1}\,,\\
\textrm{4 eqs}: \quad &&1-u_{1,3}&=u_4 \tu_4 u_{4,1} u_{4,2} u_{2,4} u_{3,4}\,, \\
\textrm{8 eqs}: \quad &&1-u_1&=\tu_2 \tu_3 \tu_4 u_{2,3} u_{3,4} u_{2,4}\,. 
\end{split}
\ea
The compatibility degrees in \eqref{D4} are obtained from walks as discussed above. Let's show an example of the walk which gives $b||a$ with $b=1\,2$ (when 3 sources can be mutated at some steps we do all of them together):

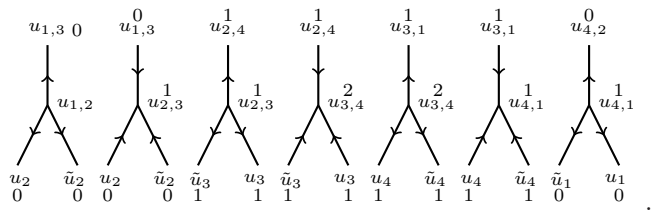
\vspace{-1.5em}
\begin{center}
\begin{tikzpicture}[scale=0.4]
\begin{scope}[thick,decoration={
            markings,
            mark=at position 0.5 with {\arrow{>}},
            }, shift={(0.5,0)},scale=2] 
            \draw[postaction={decorate}] (-2,0) -- (-1.5,-1);
            \draw[postaction={decorate}] (-2,0) -- (-2,1);
            \draw[postaction={decorate}] (-2,0) -- (-2.5,-1) ;
            
            \draw[postaction={decorate}] (-1,-1) -- (-0.5,0);
            \draw[postaction={decorate}] (-0.5,1) -- (-0.5,0);
            \draw[postaction={decorate}] (0,-1) -- (-0.5,0);
            
            \draw[postaction={decorate}] (1,0) -- (1.5,-1);
            \draw[postaction={decorate}] (1,0) -- (1,1);
            \draw[postaction={decorate}] (1,0) -- (0.5,-1) ;
            
            \draw[postaction={decorate}] (2,-1) -- (2.5,0);
            \draw[postaction={decorate}] (2.5,1) -- (2.5,0);
            \draw[postaction={decorate}] (3,-1) -- (2.5,0);
            
            \draw[postaction={decorate}] (4,0) -- (4.5,-1);
            \draw[postaction={decorate}] (4,0) -- (4,1);
            \draw[postaction={decorate}] (4,0) -- (3.5,-1) ;
            
            \draw[postaction={decorate}] (5,-1) -- (5.5,0);
            \draw[postaction={decorate}] (5.5,1) -- (5.5,0);
            \draw[postaction={decorate}] (6,-1) -- (5.5,0);
            
            \draw[postaction={decorate}] (7,0) -- (7.5,-1);
            \draw[postaction={decorate}] (7,0) -- (7,1);
            \draw[postaction={decorate}] (7,0) -- (6.5,-1) ;
        \end{scope}

\node at (-3.5,2.5) {\scriptsize $u_{1,3}$};
\node[right] at (-3,2.5) {\scriptsize $0$};
\node[right] at (-3.5,0) {\scriptsize $u_{1,2}$};
\node at (-4.4,-2.5) {\scriptsize $u_{2}$};
\node at (-2.6,-2.45) {\scriptsize $\tilde u_{2}$};
\node at (-2.5,-3) {\scriptsize $0$};
\node at (-4.5,-3) {\scriptsize $0$};

\node at (-0.5,2.5) {\scriptsize $u_{1,3}$};
\node[right] at (0,0.5) {\scriptsize $1$};
\node[right] at (-0.5,0) {\scriptsize $u_{2,3}$};
\node at (-1.4,-2.5) {\scriptsize $u_{2}$};
\node at (0.4,-2.45) {\scriptsize $\tilde u_{2}$};
\node at (0.5,-3) {\scriptsize $0$};
\node at (-0.5,3) {\scriptsize $0$};
\node at (-1.5,-3) {\scriptsize $0$};

\node at (2.5,2.5) {\scriptsize $u_{2,4}$};
\node[right] at (3,0.5) {\scriptsize $1$};
\node[right] at (2.5,0) {\scriptsize $u_{2,3}$};
\node at (1.6,-2.5) {\scriptsize $\tilde u_3$};
\node at (3.4,-2.45) {\scriptsize $u_3$};
\node at (3.5,-3) {\scriptsize $1$};
\node at (2.5,3) {\scriptsize $1$};
\node at (1.5,-3) {\scriptsize $1$};

\node at (5.5,2.5) {\scriptsize $u_{2,4}$};
\node[right] at (6,0.5) {\scriptsize $2$};
\node[right] at (5.5,0) {\scriptsize $u_{3,4}$};
\node at (4.6,-2.5) {\scriptsize $\tilde u_3$};
\node at (6.4,-2.45) {\scriptsize $u_3$};
\node at (6.5,-3) {\scriptsize $1$};
\node at (5.5,3) {\scriptsize $1$};
\node at (4.5,-3) {\scriptsize $1$};

\node at (8.5,2.5) {\scriptsize $u_{3,1}$};
\node[right] at (9,0.5) {\scriptsize $2$};
\node[right] at (8.5,0) {\scriptsize $u_{3,4}$};
\node at (7.6,-2.5) {\scriptsize $u_4$};
\node at (9.4,-2.45) {\scriptsize $\tilde u_4$};
\node at (9.5,-3) {\scriptsize $1$};
\node at (8.5,3) {\scriptsize $1$};
\node at (7.5,-3) {\scriptsize $1$};

\node at (11.5,2.5) {\scriptsize $u_{3,1}$};
\node[right] at (12,0.5) {\scriptsize $1$};
\node[right] at (11.5,0) {\scriptsize $u_{4,1}$};
\node at (10.6,-2.5) {\scriptsize $u_4$};
\node at (12.4,-2.45) {\scriptsize $\tilde u_4$};
\node at (12.5,-3) {\scriptsize $1$};
\node at (11.5,3) {\scriptsize $1$};
\node at (10.5,-3) {\scriptsize $1$};

\node at (14.5,2.5) {\scriptsize $u_{4,2}$};
\node[right] at (15,0.5) {\scriptsize $1$};
\node[right] at (14.5,0) {\scriptsize $u_{4,1}$};
\node at (13.6,-2.5) {\scriptsize $\tilde u_1$};
\node at (15.4,-2.45) {\scriptsize $ u_1$};
\node at (15.5,-3) {\scriptsize $0$};
\node at (14.5,3) {\scriptsize $0$};
\node at (13.5,-3) {\scriptsize $0$};
\end{tikzpicture}.
\end{center}

The non-simply-laced Dynkin diagrams can be obtained from simply-laced ones by {\it folding}.  Type ${\cal B}_{n{-}1}$ can be obtained from type ${\cal D}_n$ by identifying $u_i$ and $\tilde{u}_i$ for $i=1,\cdots, n$, thus we can obtain $N=n^2{-}n$ $u$'s and equations for ${\cal B}_{n{-}1}$ (here $b|| a$ is not necessarily symmetric)~\footnote{Note that the arrowheads here are cluster-theoretic, parallelling those in the simply-laced case, and should not be confused with the Lie-theoretic arrowheads which are sometimes drawn in non-simply laced types. 
}. For example, for ${\cal B}_3$ we have $12$ $u$-equations:
\ba
\begin{split}\label{globalb3}
1-u_{1,2}&=u_{3}^{2}u_{4}^{2}u_{4,1}u_{3,1}u_{3,4}^{2}u_{2,4}u_{2,3},\\
1-u_{1,3}&=u_{4}^{2}u_{4,2}u_{4,1}u_{3,4}u_{2,4},\\
1-u_{1}&=u_{2}u_{3}u_{4}u_{2,3}u_{3,4}u_{2,4}\,,
\end{split}
\ea
plus cyclic rotations. Similarly ${\cal C}_{n{-}1}$ can be obtained from ${\cal A}_{2n{-}3}$, which corresponds to triangulating a $2n$-gon, by identifying $u_{i,j}=u_{i{+}n, j{+}n}$ (the cluster seeds correspond to centrally-symmetric triangulations). Combinatorially, it is also an $(n{-}1)$-dim cyclohedron, but its $u$-equations, obtained from folding those of ${\cal A}_{2n{-}1}$, are different from those of ${\cal B}_{n{-}1}$ since some compatibility degrees differ, except for the case ${\cal B}_2={\cal C}_2$.

Finally let's give the only two 2d cases beyond ${\cal A}_2$: hexagon, ${\cal B}_2={\cal C}_2$, and octagon, ${\cal G}_2$. From identification of ${\cal A}_3$, and denoting the variables cyclically as $u_1, v_1, u_2, v_2, u_3, v_3$, the $u$ equations for ${\cal B}_2={\cal C}_2$ read
\be\label{B2C2}
1-u_1=u_2 v_2 u_3 \,,\quad 1-v_1=v_2 u_3^2 v_3\,,
\ee
plus 2 more cyclic rotations. Here we see the walk that gives the equation for $v_1$:
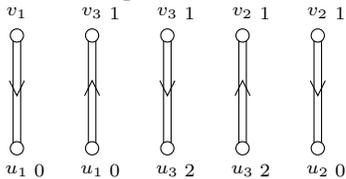
\vspace{-1em}
\begin{center}
\begin{tikzpicture}
\draw (0.95,0) -- (0.95,1.5);
\draw (1.05,0) -- (1.05,1.5);
\draw (0.9,0.9) -- (1,0.7) -- (1.1,0.9);
\node[fill=white,circle,draw=black, inner sep=0pt,minimum size=5pt] at (1,0) {};
\node at (1,-0.3) {\scriptsize $u_1$};
\node at (1.3,-0.3) {\scriptsize $0$};
\node[fill=white,circle,draw=black, inner sep=0pt,minimum size=5pt] at (1,1.5) {};
\node at (1,1.8) {\scriptsize $v_1$};

\draw (1.95,0) -- (1.95,1.5);
\draw (2.05,0) -- (2.05,1.5);
\draw (1.9,0.7) -- (2,0.9) -- (2.1,0.7);
\node[fill=white,circle,draw=black, inner sep=0pt,minimum size=5pt] at (2,0) {};
\node at (2,-0.3) {\scriptsize $u_1$};
\node at (2.3,-0.3) {\scriptsize $0$};
\node[fill=white,circle,draw=black, inner sep=0pt,minimum size=5pt] at (2,1.5) {};
\node at (2,1.8) {\scriptsize $v_3$};
\node at (2.3,1.8) {\scriptsize $1$};

\draw (2.95,0) -- (2.95,1.5);
\draw (3.05,0) -- (3.05,1.5);
\draw (2.9,0.9) -- (3,0.7) -- (3.1,0.9);
\node[fill=white,circle,draw=black, inner sep=0pt,minimum size=5pt] at (3,0) {};
\node at (3,-0.3) {\scriptsize $u_3$};
\node at (3.3,-0.3) {\scriptsize $2$};
\node[fill=white,circle,draw=black, inner sep=0pt,minimum size=5pt] at (3,1.5) {};
\node at (3,1.8) {\scriptsize $v_3$};
\node at (3.3,1.8) {\scriptsize $1$};

\draw (3.95,0) -- (3.95,1.5);
\draw (4.05,0) -- (4.05,1.5);
\draw (3.9,0.7) -- (4,0.9) -- (4.1,0.7);
\node[fill=white,circle,draw=black, inner sep=0pt,minimum size=5pt] at (4,0) {};
\node at (4,-0.3) {\scriptsize $u_3$};
\node at (4.3,-0.3) {\scriptsize $2$};
\node[fill=white,circle,draw=black, inner sep=0pt,minimum size=5pt] at (4,1.5) {};
\node at (4,1.8) {\scriptsize $v_2$};
\node at (4.3,1.8) {\scriptsize $1$};

\draw (4.95,0) -- (4.95,1.5);
\draw (5.05,0) -- (5.05,1.5);
\draw (4.9,0.9) -- (5,0.7) -- (5.1,0.9);
\node[fill=white,circle,draw=black, inner sep=0pt,minimum size=5pt] at (5,0) {};
\node at (5,-0.3) {\scriptsize $u_2$};
\node at (5.3,-0.3) {\scriptsize $0$};
\node[fill=white,circle,draw=black, inner sep=0pt,minimum size=5pt] at (5,1.5) {};
\node at (5,1.8) {\scriptsize $v_2$};
\node at (5.3,1.8) {\scriptsize $1$};

\end{tikzpicture}
\end{center}

From identification of ${\cal B}_3$ and renaming variables as $u_1, v_1,\cdots, u_4, v_4$, we have 8 $u$ equations for ${\cal G}_2$ (below we show the walk for $v_1$), which are cyclic rotations of
\be\label{G2}
1-u_1=u_2 v_2 u_3^2  v_3 u_4 \,, \quad 1-v_1=v_2 u_3^3 v_3^2 u_4^3 v_4  \,.
\ee
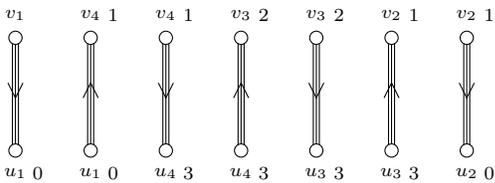
\vspace{-1.5em}
\begin{center}
\begin{tikzpicture}
\draw (0.96,0) -- (0.96,1.5);
\draw (1,0) -- (1,1.5);
\draw (1.04,0) -- (1.04,1.5);
\draw (0.9,0.9) -- (1,0.7) -- (1.1,0.9);
\node[fill=white,circle,draw=black, inner sep=0pt,minimum size=5pt] at (1,0) {};
\node at (1,-0.3) {\scriptsize $u_1$};
\node at (1.3,-0.3) {\scriptsize $0$};
\node[fill=white,circle,draw=black, inner sep=0pt,minimum size=5pt] at (1,1.5) {};
\node at (1,1.8) {\scriptsize $v_1$};

\draw (1.96,0) -- (1.96,1.5);
\draw (2,0) -- (2,1.5);
\draw (2.04,0) -- (2.04,1.5);
\draw (1.9,0.7) -- (2,0.9) -- (2.1,0.7);
\node[fill=white,circle,draw=black, inner sep=0pt,minimum size=5pt] at (2,0) {};
\node at (2,-0.3) {\scriptsize $u_1$};
\node at (2.3,-0.3) {\scriptsize $0$};
\node[fill=white,circle,draw=black, inner sep=0pt,minimum size=5pt] at (2,1.5) {};
\node at (2,1.8) {\scriptsize $v_4$};
\node at (2.3,1.8) {\scriptsize $1$};

\draw (2.96,0) -- (2.96,1.5);
\draw (3,0) -- (3,1.5);
\draw (3.04,0) -- (3.04,1.5);
\draw (2.9,0.9) -- (3,0.7) -- (3.1,0.9);
\node[fill=white,circle,draw=black, inner sep=0pt,minimum size=5pt] at (3,0) {};
\node at (3,-0.3) {\scriptsize $u_4$};
\node at (3.3,-0.3) {\scriptsize $3$};
\node[fill=white,circle,draw=black, inner sep=0pt,minimum size=5pt] at (3,1.5) {};
\node at (3,1.8) {\scriptsize $v_4$};
\node at (3.3,1.8) {\scriptsize $1$};

\draw (3.96,0) -- (3.96,1.5);
\draw (4,0) -- (4,1.5);
\draw (4.04,0) -- (4.04,1.5);
\draw (3.9,0.7) -- (4,0.9) -- (4.1,0.7);
\node[fill=white,circle,draw=black, inner sep=0pt,minimum size=5pt] at (4,0) {};
\node at (4,-0.3) {\scriptsize $u_4$};
\node at (4.3,-0.3) {\scriptsize $3$};
\node[fill=white,circle,draw=black, inner sep=0pt,minimum size=5pt] at (4,1.5) {};
\node at (4,1.8) {\scriptsize $v_3$};
\node at (4.3,1.8) {\scriptsize $2$};

\draw (4.96,0) -- (4.96,1.5);
\draw (5,0) -- (5,1.5);
\draw (5.04,0) -- (5.04,1.5);
\draw (4.9,0.9) -- (5,0.7) -- (5.1,0.9);
\node[fill=white,circle,draw=black, inner sep=0pt,minimum size=5pt] at (5,0) {};
\node at (5,-0.3) {\scriptsize $u_3$};
\node at (5.3,-0.3) {\scriptsize $3$};
\node[fill=white,circle,draw=black, inner sep=0pt,minimum size=5pt] at (5,1.5) {};
\node at (5,1.8) {\scriptsize $v_3$};
\node at (5.3,1.8) {\scriptsize $2$};

\draw (5.96,0) -- (5.96,1.5);
\draw (6,0) -- (6,1.5);
\draw (6.04,0) -- (6.04,1.5);
\draw (5.9,0.7) -- (6,0.9) -- (6.1,0.7);
\node[fill=white,circle,draw=black, inner sep=0pt,minimum size=5pt] at (6,0) {};
\node at (6,-0.3) {\scriptsize $u_3$};
\node at (6.3,-0.3) {\scriptsize $3$};
\node[fill=white,circle,draw=black, inner sep=0pt,minimum size=5pt] at (6,1.5) {};
\node at (6,1.8) {\scriptsize $v_2$};
\node at (6.3,1.8) {\scriptsize $1$};

\draw (6.96,0) -- (6.96,1.5);
\draw (7,0) -- (7,1.5);
\draw (7.04,0) -- (7.04,1.5);
\draw (6.9,0.9) -- (7,0.7) -- (7.1,0.9);
\node[fill=white,circle,draw=black, inner sep=0pt,minimum size=5pt] at (7,0) {};
\node at (7,-0.3) {\scriptsize $u_2$};
\node at (7.3,-0.3) {\scriptsize $0$};
\node[fill=white,circle,draw=black, inner sep=0pt,minimum size=5pt] at (7,1.5) {};
\node at (7,1.8) {\scriptsize $v_2$};
\node at (7.3,1.8) {\scriptsize $1$};

\end{tikzpicture}
\end{center}
Finally, we note that ${\cal F}_4$ can be obtained by folding ${\cal E}_6$.

\paragraph*{Cluster configuration space.} The first observation about the $u$-equations for a finite type ${\Phi}$ is that only $N{-}n$ of them are independent, thus the solution space, denote as $U(\Phi)$, is $n$ dimensional. Purely algebraically, the $U$ space has the same boundary structure as the generalized associahedron: as $u_a \to 0$, all incompatible $u_b \to 1$ (those with $b || a>0$), and the boundary has the same structure as the corresponding facet of the generalized associahedron. Moreover, if we impose that all $u$'s are positive, we have the {\it positive part} with all $N$ $0<u_a<1$, which cut out a``curvy" generalized associahedron of type $\Phi$. For example, $U^+({\cal B}), U^+({\cal C})$ give ``curvy" cyclohedra whose facets are of the shape ${\cal A} \times {\cal B}$ (or ${\cal A} \times {\cal C}$); $U^+({\cal D}_4)$ is a curvy polytope with 12 ${\cal A}_3 \sim {\cal D}_3$ facets and 4 ${\cal A}_1^3$ (cubes), and in total $50$ vertices (one for each seed when $4$ compatible $u$'s approach zero). 

We remark that the $u$-equations for any finite type can be derived from a set of equations we call {\it local} $u$-equations, which are associated with the walks. For any walk with $N{-}n$ steps, at each step we write 
\be\label{localu}
\frac{1-u_v}{u_v} \frac{1-u_{v'}}{u_{v'}}=\prod_{w \leftarrow v} (1-u_w)\,,
\ee
for mutating a source $v$ to $v'$. 
By considering different walks, we find that there are $N$ such local $u$ equations \eqref{localu} with rank $N{-}n$; we claim that they are equivalent to \eqref{ueqs}. For example, one can show that \eqref{global} are equivalent to the following $\frac{n(n{-}3)}{2}$ equations for ${\cal A}_{n{-}3}$:
\be
\frac{1-u_{i,j}}{u_{i,j}} \frac{1-u_{i{+}1, j{+}1}}{u_{i{+}1, j{+}1}}=(1-u_{i,j{+}1})(1-u_{i{+}1, j})\,.
\ee
For ${\cal B}_2={\cal C}_2$ and ${\cal G}_2$, \eqref{B2C2} and \eqref{G2} are equivalent to
\be
\frac{1-u_i}{u_i}\frac{1-u_{i{+}1}}{u_{i{+}1}}=1-v_i\,,\quad \frac{1-v_i}{v_i}\frac{1-v_{i{+}1}}{v_{i{+}1}}=(1-u_{i{+}1})^p\,,\nonumber
\ee
for $p=2,3$, respectively (note we have $i=1,2,\cdots, p{+}1$). With the identification $Y:=\frac{u}{1-u}$, the local $u$ equations agree with recurrence relations for these $Y$ variables in the Y system for any finite type cluster algebra~\cite{fomin2003systems, ZAMOLODCHIKOV1991391}. 

From \eqref{localu} we see that only for an acyclic quiver, all $N$ $u$-variables can be solved {\it rationally} using its $n$ variables. One can then show that the form $\prod_\alpha d\log \frac{u_\alpha}{1-u_\alpha}$ is identical (up to a sign) for all such seeds, which in turn gives the canonical cluster form $\Omega(U^+)$ with logarithmic singularities on the boundaries of the $U$ space.

\section{Open cluster string integrals}

Based on any binary positive geometry, it is natural to write down stringy integrals generalizing usual string amplitudes. To write down the canonical form of $U^+$, we apply exactly the same prescription as in the type ${\cal A}$ case. Pick any seed with an acyclic quiver which has $u_\alpha$ with $\alpha$ varying over the $n$ nodes, {\it e.g.} for ${\cal D}_n$ we can choose $u_1, u_1', u_{12}, \cdots, u_{1, n{-}1}$; we find a diffeomorphism from $\{ 0<u_\alpha<1\}$ to $U_n^+(\Phi)$, thus we have a (trivial) pushforward formula for $\Omega(U^+({\Phi}))$ as in \eqref{canonicalform}. 
The {\bf open cluster string integral} over $U^+(\Phi)$ is defined as:
\be
{\cal I}^{\Phi} (\{X\}):=(\alpha')^n \int_{U^+(\Phi)}~\prod_{\alpha}^n d\log \frac{u_\alpha}{1-u_\alpha}~\prod_{a}^N u_a^{\alpha' X_a}\,, \nonumber
\ee
where $X_a>0$ and we have a meromorphic function of $X$'s which is reminiscent of string amplitudes. Such integrals are exponentially suppressed as $\alpha'\to \infty$, and satisfy the analog of channel-duality and Regge behavior~\cite{Arkani-Hamed:2019mrd}. It is the stringy canonical forms of an ABHY generalized associahedron~\cite{Arkani-Hamed:2019mrd} (see sec. 9.3 for a ${\cal D}_4$ example): if we choose a positive parametrization using {\it e.g.} principal coefficients, the regulator $\prod u^X$ contains the $N{-}n$ $F$-polynomials.  As $\alpha'\to 0$, we have the ABHY polytope for the dual Dynkin diagram, which is given by the Minkowski sum of the Newton polytopes~\cite{20201}. Its canonical function gives the leading order of the integral, $\sum_{\rm seed} \prod_v \frac 1{X_v}$, where we sum over all vertices/seeds with each term given by product of $X_v$ for its $d$ facets. For ${\cal B}_{n{-}1}/{\cal C}_{n{-}1}$ and ${\cal D}_n$, it gives planar $n$-point tadpole diagrams, and one-loop planar integrand for bi-adjoint $\phi^3$-theory, respectively~\cite{Arkani-Hamed:2019vag}. The same result can be obtained from pushforward due to the ``scattering-equation map" from $U^+$ to the ABHY polytope. 

These generalized string amplitudes are very special since on any ``massless" pole, $X_a=0$, they have factorizations associated with Dynkin diagrams, at finite $\alpha'$! The argument is the same as for type ${\cal A}$: the residue at $X_a=0$ is given by the integral over the boundary with $u_a \to 0$, where both the form $\Omega$ and the regulator $\prod u^X$ ``factorize" by removing a node of the Dynkin diagram. 

For example, removing a generic node in type ${\cal D}_n$ corresponds to $X_{i,j}\to 0$, and the integral ${\cal I}({\cal D}_n)$ factorizes as ${\cal I}({\cal A}) \times {\cal I} (\cal D)$ (there are $n(n{-}3)$ such facets); removing the trivalent node corresponds to $X_{i,i{+}1}\to 0$, and the integral factorizes as ${\cal I}({\cal A}_{n{-}3}) \times {\cal I} ({\cal A}_1) \times {\cal I}({\cal A}_1)$ ($n$ such facets); for removing one of the two special nodes, $X_i\to 0$ (or $\tilde{X}_i \to 0$), the residue is given by the ``forward limit" of the $(n{+}2)$-point string integral, ${\cal I}({\cal A}_{n{-}1})$. More exotic factorizations happen for type ${\cal E}_m$ with $m=6,7,8$, {\it e.g.} for ${\cal I}({\cal E}_6)$ the polytope has $6\times 7=42$ facets: for any of the $7$ poles associated with the trivalent node, the residue factorizes as ${\cal I}({\cal A}_2) \times {\cal I}({\cal A}_2) \times {\cal I}({\cal A}_1)$ or ``5-pt" $\times$ ``5-pt" $\times $ ``4-pt", and on other facets we have ${\cal I}({\cal A}_4) \times {\cal I}({\cal A}_1)$, ${\cal I} ({\cal D}_5)$, or ${\cal I}({\cal A}_5)$. It is remarkable that these integrals contain products of string amplitudes as residues on massless poles. 

\section{Closed cluster string integrals}

Finally, let's briefly comment on extended $u$-equations, and the analogs of ``orderings" and closed-string integrals for the cluster configuration space. For any finite type $\Phi$, one has extended $u$-equations of the form
\be \label{eq:UU} 
U_I + V_I=1\,,
\ee
where $I$ indexes a {\it mutation relation}: just as for type ${\cal A}$ where \eqref{extendglobal} corresponds to a mutation from $(a c) \to (b d)$, we show in~\cite{20201} that equation \eqref{eq:UU} corresponds to mutation relations of the cluster algebra.  Here $U_I$ and $V_I$ are monomials of $u$ variables determined by a mutation, and for a special mutation in our walk, it reduces to one of the usual $u$-equations \eqref{ueqs} with $U_I=u_a$ and $V_I=\prod_b u_b^{b ||  a}$. 

We conjecture that the ``orderings", or connected components of $U(\mathbb{R})$ (with boundaries removed), are in bijection with the sign patterns consistent with all the extended $u$-equations \eqref{eq:UU}. A new phenomenon is that in general these regions can have shapes different from the original positive part. For $U({\cal C}_n)$ we find that the real space is tiled by components that are ``curvy" cyclohedra and associahedra, {\it e.g.} for ${\cal B}_2={\cal C}_2$ we find $4$ hexagons and $12$ pentagons tiling the space $U(\mathbb{R})$. 

Based on the ``orderings", one can also define closed cluster string integrals over $U(\mathbb{C})$ for finite-type cluster algebras, with canonical forms and regulators for two ``orderings" $\Omega(\alpha) \prod u^X~\left(\Omega(\beta) \prod u^X \right)^*$ (and similarly open-stringy integrals with forms for two components). Such a complex integral is well-defined  if we can write $u$'s in ordering $\beta$ as monomials of $u$'s in ordering $\alpha$, so that exponents of any $u$ and $u^*$ differ only by integer shifts. We leave detailed discussion of the integrals and possible physical meaning to future work. 

\section{Point Counts}

To find the connected components of $U(\Phi)({\mathbb R})$ for any finite type, it is useful to study various topological properties of the $U$ space~\cite{20201}. The latter can be done beautifully by counting the number of points in $U$ over a finite field $\mathbb{F}_q$, ${\cal N}(q)$, for any prime number $q$ (excluding some bad prime cases). It is remarkable that ${\cal N}(q)$ is a {\it polynomial} of $q$ for type ${\cal A}$, ${\cal B}$ and ${\cal C}$, and it's a quasi-polynomial for cases including ${\cal D}_4$, ${\cal G}_2$ {\it etc.}  For these cases, $|{\cal N}(q{=}{-}1)|$ and $|{\cal N}(q{=}1)|$ immediately gives the number of connected components of $U(\mathbb{R})$, and the Euler characteristic of $U(\mathbb{C})$ (which is the number of saddle points/independent integral functions), respectively. Moreover, the coefficient of each order in $q$ is the dimension of the corresponding (co-)homology group, and in particular $|{\cal N}(q{=}0)|$ gives the number of linearly independent top-dimensional $d\log$ forms. 

Let's spell out the counting for simple cases. For ${\cal A}_{n{-}3}$, the point count gives $(q{-}2)(q{-}3)\cdots (q{-}n{+}2)$, and we recover $(n{-}1)!/2$, $(n{-}2)!$ and $(n{-}3)!$ as mentioned above. For $U({\cal B}_n)$ we find ${\cal N}(q)=(q{-}n{-}1)^n$, thus the number of components, top-dim $d\log$ forms and saddle points are $(n{+}2)^n$, $(n{+}1)^n$, and $n^n$, respectively. For $U({\cal C}_n)$, ${\cal N}(q)= (q{-}n{-}1)(q{-}3)(q{-}5)\cdots (q{-}2n{+}1)$, thus these numbers are $(2n)!!(n{+}2)$, $(2n{+}1)!! (n{+}1)$ and$(2n)!!/2$. 

Let's include two examples with quasi-polynomial counting. For ${\cal G}_2$, we find ${\cal N}(q)=(q{-}4)^2+4 \delta_q$ where $\delta_q=0$ if $q=2 \mod 3$ and $1$ if $q=1 \mod 3$; substituting $q=-1$ and $q=1$ in both cases gives the correct counting: $25$ connected components and $13$ saddle points.  For ${\cal D}_4$, ${\cal N}(q) =q^4-16 q^3+ 93 q^2-231 q+206+ 2 \delta_q$, and indeed we find $547$ components and $55$ saddle points.

\section{Outlook}

There are a large number of obvious mathematical questions associated with the binary geometries we have introduced in this letter, some of which will be taken up in \cite{20201}. But the most urgent and interesting question is a physical one: is there any natural physical meaning to the cluster string integrals beyond the usual string amplitudes associated with type ${\cal A}$? We know that these objects strikingly generalize the remarkable factorization properties of string amplitudes, and we also know that in the $\alpha^\prime \to 0$ limit, at least for the classical ${\cal B}_n, {\cal C}_n, {\cal D}_n$ types, they reduce to field theory integrands at one-loop. Are they somehow related to real string amplitudes at one-loop? If so the string loop amplitudes are clearly being represented in a completely different way. If not, what is the physical interpretation of these functions? 

Let us close by remarking that the existence of the $u$-equations giving a binary realization of the compatibiltiy for cluster polytopes is rather miraculous. Of course we can write down such $u$-equations for any $n$-dimensional polytope: for any facet $f$ and all facets $f'$ that are not adjacent to $f$, we ask $u_f + \prod_f' u_f'=1$. But naively there are as many equations as unknowns here, and we would find only a discrete set of solutions, rather than an $n$-dimensional space! It is highly non-trivial that these equations are consistent with each other and give an $n$-dimensional solution space. Indeed, working with the simplest case of $n{=}2$ dimensional polygons, one can attempt to write e.g. the $u$-equations with all unit exponents for any $N$-gon, and remarkably find that they are only consistent for $N=4,5$, which correspond to ${\cal A}_1 \times {\cal A}_1$, ${\cal A}_2$. By allowing non-trivial exponents we then find $N=6,8$ corresponding to ${\cal B}_2={\cal C}_2$ and ${\cal G}_2$. Preliminary investigations~\cite{20202} suggest that one way to obtain such binary geometries is to consider ``degenerations" of generalized associahedra with certain $c$'s set to zero, including products of lower-dim ones but many more non-trivial cases. 
It is fascinating that this simple and natural mathematical question, about binary realizations of the combinatorics of polytopal geometry, can be realized by incredibly special polytopes with a factorizing boundary structures, including the ones associated with unitary particle scattering processes in spacetime. 



\begin{acknowledgments}
We thank Francis Brown, Zhenjie Li, Lecheng Ren, Giulio Salvatori, Chi Zhang for discussions. S.H. and H.T. would like to thank Institute for Advanced Study, Princeton and Center for Mathematical Sciences and Applications, Harvard, for hospitality during various stages of the work. S.H. was supported in part by the Thousand Young Talents program, 
and NSF of China under Grant No.  11947302 and 11935013. T.L. was supported by NSF DMS-1464693 and by a von Neumann Fellowship from the Institute for Advanced Study. H.T. was partially supported by an NSERC Discovery Grant and the Canada Research Chairs program.
\end{acknowledgments}

\bibliography{ref}
\end{document}